\documentclass[12pt]{article}

\usepackage{amssymb, amsmath}
\usepackage{graphicx}
\usepackage{cite}
\def\be{\begin{equation}}
\def\ee{\end{equation}}
\def\bea{\begin{eqnarray}}
\def\eea{\end{eqnarray}}
\newtheorem{lemma}{Lemma}

\title{ {\bf Two remarks on near-horizon geometries}}

\author{James Lucietti \\  \small \sl  School of Mathematics and Maxwell Institute for Mathematical Sciences, \\ \small \sl  University of Edinburgh, King's Buildings, Edinburgh, EH9 3JZ, UK \\ \small \sl 
E-mail: j.lucietti@ed.ac.uk}

\date{}

\begin{document}

\maketitle

\begin{abstract}
We show that any extreme black hole with an orthogonally transitive abelian isometry group has a near-horizon geometry with enhanced symmetry.
We also point out a simple proof  of the horizon topology theorem of Galloway and Schoen for degenerate horizons.
\end{abstract}

\section{Introduction}
All known extreme black holes possess a near-horizon geometry with an AdS$_2$-symmetry. This has played an important role in developing a quantum description for various extreme black holes.\footnote{There is a large and ongoing literature on this, see e.g.~\cite{Strominger:1998yg} and more recently~\cite{Sen:2008yk}.} For $D=4,5$ the existence of this symmetry was proved~\cite{Kunduri:2007vf} for extreme black hole solutions to a wide class of theories of Einstein gravity coupled to an arbitrary number of Maxwell and neutral scalar fields, by assuming a $U(1)^{D-3}$ rotational symmetry (and non-toroidal horizon topology). For $D>5$ the near-horizon symmetry enhancement phenomenon still occurs for the known solutions, such as the Myers-Perry black holes, but a general understanding is lacking unless one assumes a cohomogeneity-1 rotational symmetry~\cite{Figueras:2008qh}. 

On the other hand, a crucial step in the electro-vacuum $D=4$ black hole uniqueness theorem, is that for stationary and axisymmetric spacetimes the Einstein equations imply orthogonal transitivity\footnote{An isometry group whose surfaces of transitivity are $p<D$ dimensional is said to be orthogonally transitive if there exists $D-p$ dimensional surfaces orthogonal to the surfaces of transitivity at every point.} of the isometry group, see e.g.~\cite{Chrusciel:2012jk}.  This result generalises to $D 
\geq 5$ by assuming an $\mathbb{R}\times U(1)^{D-3}$ isometry group (for vacuum see~\cite{Emparan:2001wk})\footnote{This assumes that some combination of the rotational Killing field has a fixed point.}, which for $D=5$ captures all known explicit solutions. For $D>5$, the known higher cohomogeneity black holes, such as the Myers-Perry solutions,  still possess orthogonal transitivity; however, we lack any general understanding of its relation to the Einstein equations (indeed there may be none). 

Our first remark, and main result, relates these two phenomena. We show that any extreme black hole with an orthogonally transitive abelian isometry group, has a near-horizon geometry with  an enhanced symmetry. Furthermore, in many theories of interest this symmetry must be that of AdS$_2$ (specifically those where the strong energy condition holds). Our precise results are stated and proven in Sec. \ref{sec:symmetry}.

Our second, independent, remark is that Galloway and Schoen's horizon topology theorem~\cite{Galloway:2005mf} has a simple proof for degenerate horizons, which we present in Sec. \ref{sec:topology}.

\section{Near-horizon symmetry}
\label{sec:symmetry}

Consider a $D$-dimensional spacetime containing a smooth degenerate Killing horizon  $\mathcal{N}$  of a complete future-directed Killing field $K$, with a cross-section $H$ (i.e. a $D-2$ dimensional spacelike submanifold of $\mathcal{N}$ which is intersected once by each orbit of $K$). Let $U$ be the unique past-directed vector field that is tangent to null geodesics which are orthogonal to $H$ and satisfy $K \cdot U=1$.

In the neighbourhood of any such degenerate horizon one can define a scaling limit which focuses on the geometry near $\mathcal{N}$~\cite{Kunduri:2007vf}. The resulting near-horizon geometry, in Gaussian null coordinates, is
\be
ds^2= 2\, dv \left( dr +r h_a(x) dx^a + \tfrac{1}{2} r^2F(x)dv \right) +\gamma_{ab}(x) dx^a dx^b  \label{NHG}
\ee
where $K=\partial / \partial v$,  $U=\partial /\partial r$, $\mathcal{N}= \{ r=0\}$ and $(x^a)$, $a=1,\dots, D-2$, are coordinates on $H=\{ r=0, v=\text{const} \}$. Here $\gamma_{ab}$ is the induced (Riemannian) metric on $H$,  $h_a$ is 1-form on $H$ and $F$ is a function on $H$.

We are now ready to present our first and main result.
\begin{lemma} \label{lemma1}
Consider a spacetime containing a degenerate horizon as above. Assume the spacetime has an orthogonally transitive isometry group $\mathbb{R}\times U(1)^N$, where $1\leq N\leq D-3$ and $\mathbb{R}$ is generated by $K$, such that the surfaces orthogonal to the surfaces of transitivity are simply connected. Then, the near-horizon geometry has an isometry group $G \times U(1)^N$, where $G$ is the isometry group of AdS$_2$,  dS$_2$ or 2d Minkowski space.
\end{lemma}

\noindent {\it Proof}: Let $m_I$, for $I=1,\dots, N$, denote the Killing fields which generate the assumed $U(1)^N$ isometry. This isometry group leaves $\mathcal{N}$ invariant (since $m_I$ and $K$ commute) and hence the $m_I$ must be tangent to the horizon. In particular, since the $m_I$ have closed orbits they must tangent to $H$. Now, define the following set of differential forms, for $I=1, \dots, N$, 
\be
 \Omega =K \wedge m_1 \wedge \dots \wedge m_N \wedge dK \qquad \text{and} \qquad \omega_I = K \wedge m_1 \wedge \dots \wedge m_N \wedge dm_I  \; ,
\ee
where (by abuse of notation) the 1-forms  $K_\mu = g_{\mu\nu} K^\nu$ and $(m_I)_{\mu}=g_{\mu\nu} (m_I)^\nu$. Frobenius' integrability theorem states that the isometry group generated by the Killing vector fields $K$ and $m_I$ is orthogonally transitive if and only if $\Omega=0$ and $\omega_I=0$ for all $I=1, \dots, N$. 

It follows that  $U(1)^N$ is orthogonally transitive on $H$, as we now show. First note that the $m_I$ generate a $U(1)^N$ isometry of $\gamma_{ab}$ when restricted to $H$. Denoting the restriction of the dual 1-forms by $\hat{m}_I= m_I|_H$, it is easily checked that $i_U\omega_I|_{H}=  \hat{m}_1 \wedge \cdots \wedge \hat{m}_N \wedge d\hat{m}_I \equiv \hat{\omega}_I$. Therefore $\omega_I=0$ implies $\hat{\omega}_I=0$, which by Frobenius' theorem shows that the isometry $U(1)^N$ is orthogonally transitive on $H$, as claimed.\footnote{Note that the arguments so far are valid for the full spacetime (i.e. not just its near-horizon geometry) and also for non-degenerate horizons.}

The near-horizon geometry (\ref{NHG}) inherits orthogonal transitivity of $\mathbb{R} \times U(1)^N$ from the full spacetime. Let $(\phi^I)$ be coordinates on the surfaces of transitivity of $U(1)^N$ adapted to the Killing fields $m_I=\partial / \partial \phi^I$, and let $(y^m)$ be coordinates on the surfaces $\Sigma$ in $H$ orthogonal to the surfaces of transitivity. In these coordinates the near-horizon geometry (\ref{NHG}) is given by
\bea
\gamma_{ab}dx^adx^b = \gamma_{mn}(y) dy^m dy^n +\gamma_{IJ}(y) d\phi^Id\phi^J  \; ,
\eea
$h_adx^a = h_m(y) dy^m + h_I(y) d\phi^I$ and $F=F(y)$, so all dependence is on the coordinates $(y^m)$.   A tedious calculation shows that $\Omega=0$ is equivalent to
\be
\partial_m F-Fh_m -h^I\partial_m h_I= 0 \; \quad \text{and} \qquad \partial_{[n}h_{m]}=0 \; ,  \label{Feq}
\ee
whereas $\omega_I=0$ is equivalent to
\be
\partial_m h_{I}-h^J\partial_m \gamma_{IJ} -h_Ih_m=0 \; ,   \label{heq}
\ee
where $h^I= \gamma^{IJ}h_J$ and $\gamma^{IJ}$ is the inverse of $\gamma_{IJ}$.\footnote{The identities $m_1\wedge \dots \wedge m_N= \gamma \; d\phi^1 \wedge \dots  \wedge d\phi^N  +(-1)^{N-1} \sqrt{\gamma} \; rdv \wedge \star_\gamma h_I d\phi^I$ and $(\star_\gamma d\phi^K )\wedge d\phi^J = \sqrt{\gamma} \gamma^{KJ} d\phi^1\wedge \dots \wedge d\phi^N$, 
where $\gamma=\det \gamma_{IJ}$ and $\star_\gamma$ is the Hodge dual with respect to $\gamma_{IJ}$, were useful in performing these calculations.}

We may solve the PDEs (\ref{Feq}) and (\ref{heq}) as follows. The second equation in (\ref{Feq}) implies that locally there exists a function $\lambda$ such that $h_m=\partial_m \lambda$;  simply connectedness of $\Sigma$ implies that $\lambda$ is globally defined. Substituting into equation (\ref{heq}) then implies
\be
h^I= e^{\lambda} k^I
\ee
where $k^I$ are constants. Finally, plugging into the first equation in (\ref{Feq}) gives
\be
F= A_0 e^{\lambda}+ \gamma_{IJ}k^Ik^J e^{2\lambda}
\ee
where $A_0$ is a constant. We may now substitute these expressions into the full near-horizon geometry. Changing the coordinate $r\to e^{-\lambda} r$ results in
\be
ds^2= e^{-\lambda}[ A_0 r^2 dv^2 +2dvdr] + \gamma_{IJ}( d\phi^I+k^I rdv)(d\phi^J+k^Jrdv) +\gamma_{mn} dy^m dy^n  \;.
\ee
The metric in the square brackets is that of AdS$_2$, dS$_2$ or 2d Minkowski space, depending on $A_0$. Any isometry of these 2d spaces maps $rdv\to rdv +d\psi$ for some function $\psi$; if we also map $\phi^I \to \phi^I-k^I \psi$ we get  the claimed isometry of the near-horizon geometry.  \\

Let us emphasise that the above result does not assume any field equations and is valid in any theory of gravity (thus including higher derivative theories). By assuming a certain energy condition we can refine our statement as follows.
\begin{lemma} 
Consider the setup in Lemma \ref{lemma1}.  Further, suppose  $R_{\mu \nu} K^\mu U^\nu |_H \leq 0$ holds and $H$ is compact. Then a non-trivial near-horizon geometry must have AdS$_2$-symmetry.
\end{lemma}

\noindent {\it Proof:} The $vr$ component of the Ricci tensor of the near-horizon geometry (\ref{NHG}) is given by
\be
F = \tfrac{1}{2} h_ah^a - \tfrac{1}{2}\nabla_a h^a+R_{\mu \nu} K^\mu U^\nu |_H
\ee
where $\nabla_a$ is the metric connection of $\gamma_{ab}$,  see e.g.~\cite{Kunduri:2008rs}. Note the same equation arises by restricting $R_{\mu\nu} K^\mu U^\nu$ to the horizon for the full spacetime. In our case this reduces to
\be
A_0 =-\tfrac{1}{2} e^{\lambda} \gamma_{IJ}k^I k^J + \tfrac{1}{2} \nabla^2 e^{-\lambda} + e^{-\lambda} R_{\mu \nu} K^\mu U^\nu|_H  \; .
\ee
Integrating this over compact $H$ and using the stated assumptions shows $A_0\leq 0$, with equality if and only if $h_a \equiv 0$ and $F \equiv 0$ (in which case (\ref{NHG}) is the trivial product $\mathbb{R}^{1,1} \times H$). Hence non-trivial near-horizon geometries in this class must have $A_0<0$ and so possess an AdS$_2$-symmetry. \\

We now make some comments on the above two results. 
\begin{itemize}
\item The assumption that the surfaces orthogonal to the surfaces of transitivity are simply connected has been established for asymptotically flat and Kaluza-Klein spacetimes in the above class satisfying the null energy condition, via topological censorship~\cite{Chrusciel:2008rh}.
\item $R_{\mu \nu} K^\mu U^\nu |_H \leq 0$ follows from the strong energy condition, as we now show. On the horizon $\mathcal{N}$, the vector field $K^\mu-U^\mu$ is (future directed) time-like, so the strong energy condition implies that $R_{\mu \nu} (K-U)^\mu (K-U)^\nu |_{\mathcal{N}} \geq 0$. Now, $R_{\mu \nu} (K-U)^\mu (K-U)^\nu |_{\mathcal{N}} = - 2R_{\mu\nu} K^\mu U^{\nu}|_{\mathcal{N}}$, since $R_{\mu\nu} K^\mu K^{\nu}|_{\mathcal{N}} =0$ for a Killing horizon and $R_{\mu\nu} U^\mu U^\nu =0$ for the near-horizon geometry (\ref{NHG}), so the claim follows.
\item For $N=D-3$, as mentioned in the introduction, orthogonal transitivity follows from Einstein's equations (assuming the rotational Killing fields possess fixed points); our results then imply the result established in~\cite{Kunduri:2007vf} by direct use of the Einstein equations. 
\item For $N<D-3$, our theorem guarantees a near-horizon  AdS$_2$-symmetry for all the known $D>5$ examples since they have an orthogonally transitive isometry group (with $N=\lfloor (D-1)/2 \rfloor$); this includes the Myers-Perry black holes and supergravity generalisations thereof. 
\item Recently, a scalar field instability was established for general (compact) extreme horizons under the {\it assumption} that $A_0$ is a non-zero constant~\cite{Lucietti:2012sf}. Our present results show that an orthogonally transitive isometry group $\mathbb{R}\times U(1)^N$ and strong energy imply $A_0<0$.
\end{itemize}

\section{Horizon topology}
\label{sec:topology}

Here we point out that the horizon topology theorem of Galloway and Schoen has a simple direct proof in the context of degenerate horizons. We will be brief since our argument is essentially the simplified proof for marginally outer trapped surfaces  presented in~\cite{Racz:2008tf}, but specialised to degenerate horizons. This results in a significant simplification which we feel is worth pointing out. For brevity, we consider the vacuum Einstein equations, although the proof generalises for matter satisfying the dominant energy condition. 

The Einstein equations $R_{\mu\nu}=\Lambda g_{\mu\nu}$ for a general near-horizon geometry (\ref{NHG}) are equivalent to the equation on $H$, see e.g.~\cite{Kunduri:2008rs},
\be
\label{Heq}
R_{ab} = \tfrac{1}{2}h_ah_b -\nabla_{(a} h_{b)} +\Lambda \gamma_{ab}   \; ,
\ee
where $R_{ab}$ is the Ricci tensor of $\gamma_{ab}$.  This equation can also be thought of as arising from the restriction of the Einstein equation to the degenerate horizon $\mathcal{N}$ for the full spacetime, i.e. {\it before} taking the near-horizon limit. 

Let us assume $H$ is an $n=D-2$ dimensional compact manifold. Then (\ref{Heq}) implies that if $\Lambda \geq 0$ then $\int_H R_{\gamma} \geq 0$. For $n=2$ we immediately recover Hawking's topology theorem,  however for $n>2$ this in itself does not provide a topological constraint. Nevertheless, for $n>2$, the Ricci scalar $R_\gamma$ arising from (\ref{Heq}) implies that if $\Lambda \geq 0$, a non-trivial near-horizon geometry must have cross-sections with a Yamabe invariant $\sigma(H)>0$, as we now show.  

The Yamabe invariant $\sigma(H)$ is defined via the Yamabe constant $Y(H, [\gamma ])$ associated to each conformal class $[\gamma]$, by $\sigma(H) =\sup_{[\gamma]} Y(H,[\gamma])$.  If we parameterise $[\gamma]$ by $\gamma'=\phi^{\frac{4}{n-2}}\gamma$ for positive $C^\infty$ functions $\phi$, then $Y(H,[\gamma])=  \inf_{\phi>0} Q_\gamma[\phi] $, where
\be
 Q_\gamma[\phi] \equiv  \frac{\int_{H} \left( \tfrac{4(n-1)}{n-2} |\nabla \phi|^2 + R_\gamma \phi^2\right) }{ \left(\int_{H} \phi^{\frac{2n}{n-2}}\right)^{\frac{n-2}{n}}}
 \ee
is the (normalised) Einstein-Hilbert action restricted to $[\gamma]$.   

For our case, the horizon equation (\ref{Heq}) implies that for any $\phi$
 \be
\label{id}
2 | \nabla \phi |^2 +R_\gamma \phi^2 = 2| D\phi |^2 -\nabla \cdot (\phi^2 h) +\Lambda n \phi^2  \; ,
\ee
where the operator $D_a \equiv \nabla_a +\frac{1}{2} h_a$. This in turn implies
\be
\label{intid}
\int_{H} \left( \tfrac{4(n-1)}{n-2} |\nabla \phi|^2 + R_{\gamma} \phi^2\right) =\int_H \left( 2 | D \phi |^2 + \tfrac{2n}{n-2} | \nabla \phi |^2 +n\Lambda \phi^2 \right)  \; .
\ee
For $\Lambda \geq 0$, we deduce $Q_\gamma[\phi]>0$ for all $\phi>0$,  unless $\Lambda=0$ and $h_a\equiv 0$.  The solution to the Yamabe problem guarantees that the infimum of $Q_\gamma[\phi]$ is always achieved: thus  $Q_\gamma[\phi]>0$ for all $\phi$ implies $Y(H,[\gamma])>0$ and so $\sigma(H)>0$.  We deduce that for non-trivial near-horizon geometries $\sigma(H)>0$, thus establishing our claim. For $\Lambda<0$ one does not get a constraint on the topology; instead, if $\sigma(H)<0$, one gets a lower bound on the area of $H$ as in~\cite{Racz:2008tf}.

In the above we have only used the trace of the horizon equation (\ref{Heq}). An interesting question, which we will leave for the future, is if the non-trace part of (\ref{Heq}) implies any further topological restrictions on $H$. \\

\noindent{\bf Acknowledgements.} The author is supported by an EPSRC Career Acceleration fellowship. I would like to thank Hari Kunduri and especially Harvey Reall for useful comments.

\end{document}